# Chapter 7
# Data Mining-based Techniques

Peggy Cellier[1], Mireille Ducasse[1], Sebastien Ferre[2], Olivier Ridoux[2], and W. Eric Wong[3*]

[1]IRISA/INSA Rennes
[2]2IRISA/Universite de Rennes
[3]Department of Computer Science, University of Texas at Dallas, Richardson, TX, USA

[*]Corresponding Author: W. Eric Wong; Email: ewong@utdallas.edu

## 7.1. Preface[1]

Software engineering processes generate a large amount of data, and several authors have advocated for the use of data mining methods to conduct fault localization. Many data mining methods with different merits exist, but the very first step is often to simply consider as data what was previously considered a mere by-product of a process. One of the first historical examples of uncovering new knowledge from pre-existing data was Florence Nightingale's (1820-1910) demonstration that soldiers died more often from bad sanitary conditions in military hospitals than from battle wounds. For her demonstration, she gathered previously ignored data and presented it in revealing graphics. This example demonstrates that data mining is itself a process with important questions, from the selection and gathering of data to the presentation of the results. In the fault localization context, the following questions emerge: which existing data can be leveraged to improve the localization? Which presentation of the results is best suited for the debugging process? In this chapter, we will illustrate the basic concepts of fault localization using a data mining technique and present three approaches to locate the faulty statements of a program. We will also address how data mining can be further applied to fault localization for GUI components.

## 7.2. Introduction[2]

As software engineering data are symbolic by nature, in this chapter, we will present fault localization using symbolic methods. Symbolic methods tend to lend themselves naturally to give explanations, and this is exactly what we are looking for in fault localization. Indeed, we prefer a system with the capacity of saying "the failure has to do with the initialization of variable $x$" to a system limited to saying "the fault is in these million lines with probability 0.527". Therefore, we will illustrate how to use two data mining techniques: association rules and formal concept analysis in fault localization.

Formal concept analysis and association rules deal with collections of objects and their features. The former extracts contextual truth, such as "in this assembly, all white-haired female wear glasses", while the latter extracts relativized truth, such as "in this assembly, carrying a briefcase increases the chance of wearing a tie". In a fault localization context, the former could say that "all failed tests call method $m$", and the latter could discover that "most failed tests call method $m$, which is very seldom called in passed tests".

Throughout this chapter, we use the Trityp program (partly given in Table 7.1) to illustrate the general method. It is a classical benchmark for test generation methods. Its specification is to classify sets of three segment lengths into four categories: scalene, isosceles, equilateral, and not a triangle, according to whether a given kind of triangle can be formed with these dimensions, or no triangle at all.

---

[1] Part of Chapter 7.1 is from Reference [6]
[2] Part of Chapter 7.2 is from Reference [6]

We use this benchmark to explain the ability of data mining process for localizing faults. We do so by introducing faults in the program in order to form slight variants, called mutants, and by testing them through a test suite [9]. The data mining process starts with the output of the tests, i.e., execution traces and pass/fail verdicts. The mutants can be found on the web[3], and we use them to illustrate the general localization method.

Table 7.2 presents the eight mutants of the Trityp program. The first mutant is used to explain in detail the method. For mutant 1, one fault has been introduced in Line 84. The condition (trityp == 2) is replaced by (trityp == 3). That fault causes a failure in two cases:

Table 7.1. Source code of the Trityp program [6].

|    | Public int Trityp() {              | 81  | if (trityp == 1 && (i+j)>k)                |
|----|------------------------------------|-----|--------------------------------------------|
| 57 | int trityp;                        | 82  | trityp = 2;                                |
| 58 | if ((i==0) \|\| (j==0) \|\| (k==0)) | 83  | else                                       |
| 59 | trityp = 4;                        | 84  | if ((trityp ==2) &&(i+k)>j)                |
| 60 | else                               | 85  | trityp = 2;                                |
| 61 | {                                  | 86  | else                                       |
| 62 | triyp = 0;                         | 87  | if ((triyp == 3) && (j+k)>j)               |
| 63 | if (i==j)                          | 88  | trityp = 2;                                |
| 64 | trityp =trityp+1;                  | 89  | else                                       |
| 65 | if (i==k)                          | 90  | trityp = 4;                                |
| 66 | trityp = trityp+2;                 | 91  | }                                          |
| 67 | if (j==k)                          | 92  | }                                          |
| 68 | trityp = trityp + 3;               | 93  | return(trityp);}                           |
| 69 | if (trityp ==0)                    | 94  | }                                          |
| 70 | {                                  | 95  |                                            |
| 71 | if ((i+j<=k) \|\|(j+k<=i) \|\| (i+k<=j=)) | 96 | public static string coversiontrityp (int i) { |
| 72 | trityp=4;                          | 97  | switch (i) {                               |
| 73 | else                               | 98  | case 1:                                    |
| 74 | trityp = 1;                        | 99  | return" scalen";                           |
| 75 | }                                  | 100 | case 2:                                    |
| 76 | else                               | 101 | return" isosceles";                        |
| 77 | {                                  | 102 | case 3:                                    |
| 78 | if (trityp>3)                      | 103 | return" equilateral";                      |
| 79 | trityp=3;                          | 104 | default:                                   |
| 80 | else                               | 105 | return" not a triangle" ;}}                |

Table 7.2. Mutants of the Trityp program [6].

| Mutant | Faulty line |
|--------|-------------|
| 1 | [84] if ((trityp == 3) && (i+k)>j) |
| 2 | [79] trityp = 0; |
| 3 | [64] trityp =i+1; |
| 4 | [87] if (trityp !=3) && (j+k)>i) |
| 5 | [65] if (i>=k) |
| 6 | [74] trityp =0; |
| 7 | [90] trityp ==3; |
| 8 | [66] trityp = trityp+20; |

(1) The first case is when trityp is equal to 2; the execution does not enter this branch and goes to the default case in lines 89 and 90.

---
[3] http://www.irisa.fr/lis/cellier/Trityp/Trityp.zip

(2) The second case is when trityp is equal to 3; the execution should go to Line 87, but due to the fault, it goes to Line 84. Indeed, if the condition (i+k>j) holds, trityp is assigned to 2. However, (i+k>j) does not always imply (j+k>i), which is the real condition to test when trityp is equal to 3. Therefore, trityp is assigned to 2, whereas 4 is expected.

The faults of mutants 2, 3, 6, and 8 are on assignments. The faults of mutants 4, 5, and 7 are on conditions. We will present more details about multiple fault situations in Chapter 8.5. In this case, we simply combine several mutations to form new mutants.

## 7.3. Formal Concept Analysis and Association Rules[4]

Formal concept analysis (FCA, [12] Erreur ! Source du renvoi introuvable.·**Erreur ! Source du renvoi introuvable.**) and association rules (AR, [[3]]) are two well-known methods for symbolic data mining. In their original inception, they both consider data in the form of an object-attribute table. In the FCA world, the table is called a formal context. In the AR world, objects are called transactions and attributes are called items, so that a line represents the items present in a given transaction. This comes from one of the first applications of AR, namely the basket analysis of retail sales. We will use both vocabularies interchangeably according to context.

*Definition 1* (Formal context and transactions). A formal context, $K$, is a triple $(O, A, d)$, where $O$ is a set of objects, $A$ is a set of attributes, and $d$ is a relation in $O \times A$. We write $(o,a) \in d$ or $oda$ equivalently. In the AR world, $A$ is called a set of items, or *itemset*, and each $\{i \in A \mid odi\}$ is the $o$th transaction.

For visualization, we will consider objects as labeling lines and attributes as labeling columns of a table. A cross sign at the intersection of line $o$ and column $a$ indicates that object $o$ has attribute $a$.

Table 7.3 is an example of context. The objects are the planets of the solar system, and the attributes are discretized properties of these planets: size, distance to the sun, and presence of moons. One can observe that all planets without moons are small, but all planets with moons except two are far from the sun. The difficulty is making similar observations in large data sets.

Both methods try to answer questions such has "which attributes entail these attributes?" or "which attributes are entailed by these attributes?". The main difference between FCA and AR is that FCA answers these questions to the letter, i.e., the mere exception to a candidate rule kills the rule, though association rules are accompanied by statistical indicators. In short, association rules can be almost true. As a consequence, rare events as well as frequent events are represented in FCA, whereas in AR, frequent events are distinguished.

Table 7.3. The Solar system context [6].

|  | Size | | | Sun Distance | | Moons | |
| --- | --- | --- | --- | --- | --- | --- | --- |
|  | small | medium | largen | near | far | with | Without |
| Mercury | × | | | × | | | × |
| Venus | × | | | × | | | × |
| Earth | × | | | × | | × | |
| Mars | × | | | × | | × | |
| Jupiter | | | × | | × | × | |
| Saturn | | | × | | × | × | |
| Uranus | | × | | | × | × | |

---

[4] Part of Chapter 7.3 is from Reference [6]

| | | | | × | × | |
|---|---|---|---|---|---|---|
| Neptune | | × | | × | × | |

### 7.3.1. Formal Concept Analysis

FCA searches for sets of objects and sets of attributes with equal significance, like *Mercury*, *Venus*, and *without moons*, and then orders the significances by their specificity.

*Definition 2* (extent /intent/formal concept) Let $K = (O, A, d)$ be a formal context. $\{o \in O | \forall a \in A . o d a\}$ is the extent of a set of attributes $A \subseteq A$. It is written as extent $(A)$. $\{a \in A | \forall o \in O . o d a\}$ is the intent of a set of objects $O \subseteq O$. It is written intent $(O)$.

A formal concept is a pair $(O, A)$ such that $A \subseteq A$, $O \subseteq O$, $intent(O) = A$ and $extent(A) = O$. $A$ is called the intent of the formal concept, and $O$ is called its extent.

Formal concepts are partially ordered by set inclusion of their intent or extent. $(O_1, A_1) < (O_2, A_2)$ iff $O_1 \subset O_2$. We say that $(O_2, A_2)$ contains $(O_1, A_1)$. In other words, $(O, A)$ forms a formal concept iff $O$ and $A$ are mutually optimal for describing each others; ie., they have same significance.

*Lemma 1* (Basic FCA results) It is worth remembering the following results:

$$extent(\emptyset) = O \text{ and } intent(\emptyset) = A$$

*extent(intent(extent(A)))* = *extent(A)* and *intent(extent(intent(O)))* = *intent(O)*. Hence, extent *circ* intent and intent *circ* extent are closure operators.

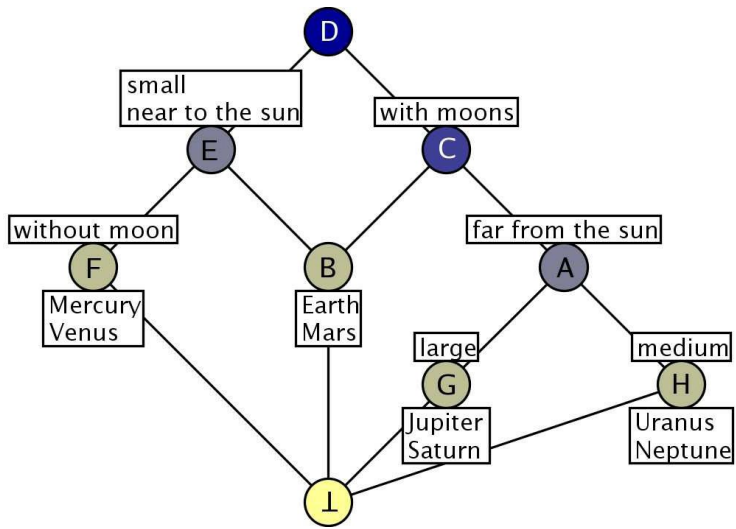

Figure 7.1. Concept lattice of the solar system context (see Table 7.1) [6].

$$(O_1, A_1) < (O_2, A_2) \text{ iff } A_1 \supset A_2$$

*(extent)intent(O)), intent(O))* is always a formal concept, and it is written as *concept(O)*. In the same way, *(extent(A), intent(extent(A)))* is always a formal concept as well, and it is written as *concept(A)*. All formal concepts can be constructed this way.

**Theorem 1** (Fundamental theorem of FCA, [16]). Given a formal context, the set of all its partially ordered formal concepts forms a lattice called a concept lattice.

Given a concept lattice, the original formal context can be reconstructed. Figure 7.1 shows the concept lattice deduced from the solar system context. It is an example of the *standard representation* of a concept lattice. In this representation, concepts are drawn as colored circles with an optional inner label that serves as a concept identifier, and 0, 1, or 2 outer labels in square boxes. Lines represent non-transitive containment; therefore, the standard representation displays a *Hasse* diagram of the lattice [26]. The figure is oriented such that higher concepts (higher in the diagram) contain lower concepts.

The upper outer label of a concept (such as large for concept $G$), when present, represents the attributes that are new to this concept intent compared with higher concepts; we call it an *attribute label*. It can be proven that if $A$ is the attribute label of concept $c$, then $A$ is the smallest set of attributes such that $c = concept(A)$. Symmetrically, the lower outer label of a concept (such as Jupiter, Saturn for concept $G$), when present, represents the objects that are new to this concept extent compared with lower concepts; we call it an object label. It can be proven that if $O$ is the object label of concept $c$, then $O$ is the smallest set of objects such that $c = concept(O)$. As a consequence, the intent of a concept is the set of all attribute labels of this concept and higher concepts, and the extent of a concept is the set of all object labels of this concept and lower concepts.

For example, the extent of concept A is {*Jupiter, Saturn, Uranus, Neptune*}, and its intent is {*far fromsun, withmoons*}. In other words, an attribute labels the highest concept to which intent it belongs, and an object labels the lowest concept to which extent it belongs.

It has been proven [12] that such a labeling where all attributes and objects are used exactly once is always possible. As a consequence, some formal concepts can be named by an attribute and/or an object; for example, concept G can be called either concept large, Jupiter, or Saturn, but others like concept D have no such names. They are merely unions or intersections of other concepts.

In the standard representation of concept lattice, "a1 entails a2" reads as an upward path from $concept(a_1)$ to $concept(a_2)$. Attributes that do not entail each other label incomparable concepts, such as attributes small and with moons. Note that there is no purely graphical way to detect that "$a_1$ nearly entails $a_2$".

The bottom concept, $\bot$, has all attributes and usually zero objects, unless some objects have all attributes. The top concept, $\top$, has all objects and usually zero attributes, unless some attributes are shared by all objects.

The worst-case time complexity of the construction of a concept lattice is exponential, but we have shown that if the size of the problem can only grow with the number of objects, i.e., the number of attributes per object is bounded, then the complexity is linear [11]. Moreover, though the mainstream interpretation of FCA is to compute the concept lattice at once and use it as a means for presenting graphically the structure of a data set, we have shown [11][21] that the concept lattice can be built and explored gradually and efficiently.

### 7.3.2. Association Rules

FCA is a crisp methodology that is sensitive to every detail of the data set. Sometimes, one may wish for a method that is more tolerant to exceptions.

**Definition 3** (Association rules), *Let K be a set of transactions, i.e., a formal context seen as a set of lines seen as itemsets. An association rule is a pair (P, C) of itemsets. It is usually written as P→ C.*

*The P part is called the premise, and the C part is the conclusion.*

Note that any *P→C* forms an association rule. It does not mean it is a relevant one. Statistical indicators give hints at the relevance of a rule.

**Definition 4** (Support/confidence/lift). *The support of a rule P→C,*

*written as* $sup(P \to C)$*, is defined as*[5]

$$extent(P \cup C)$$

*The normalized support a rule* $P \to C$ *is defined as*

$$\frac{extent(P \cup C)}{extent(\emptyset)}$$

*The confidence of rule* $P \to C$*, written* $conf(P \to C)$*, is defined as*

$$\frac{sup(P \to C)}{sup(P \to \emptyset)} = \frac{extent(P \cup C)}{extent(P)}$$

*The lift of a rule* $P \to C$*, written* $lift(P \to C)$*, is defined as*

$$\frac{conf(P \to C)}{conf(\emptyset \to C)} = \frac{sup(P \to C)}{sup(P \to \emptyset)}$$

$$sup(\emptyset \to C) = \frac{extent(P \cup C) \times extemt(\emptyset)}{extent(P) \times extent(C)}$$

Support measures the prevalence of an association rule in a data set. For example, the support of near sun with moon is 2. Normalized support measures its prevalence as a value in [0, 1], i.e., as a probability of occurrence. For example, the normalized support of near sun with moon is 2/8 = 0.25. It can be read as the probability of observing the rule in a random transaction of the context. It would seem that the greater the support, the better it is, but very often one must be satisfied with a very small support. This is because in large contexts, with many transactions and items, any given co-occurrence of several items is a rare event. Efficient algorithms exist for calculating all ARs with minimal support [2][4][24][28].

Confidence measures the "truthness" of an association rule as the ratio of the prevalence of its premise and conclusion together on the prevalence of its premise alone. Its value is in [0, 1], and for a given premise, bigger is better; in other words, it is better to have fewer exceptions to the rule considered as a logical implication. For example, the confidence of near sun with moon is 2/4 = 0.5. This can be read as the conditional probability of observing the conclusion knowing that the premise holds. However, there is no way to tell whether a confidence value is good in itself. In other words, there is no absolute threshold above which a confidence value is good.

---

[5] Where . is the cardinal of a set; how many elements it contains.

Lift also measures "truthness" of an association rule, but instead as the increase of the probability of observing the conclusion when the premise holds w.r.t. when it does not hold. In other words, it measures how the premise of a rule increases the chance of observing the conclusion. A lift value of 1 indicates that the premise and conclusion are independent. A lower value indicates that the premise repels the conclusion, and a higher value indicates that the premise attracts the conclusion. For example, the lift of near sun with moon is 0.5/0.75, which shows that the attribute near sun repels the attribute with moon; to be near the sun diminishes the probability of having a moon. The rule near sun without moon has a support value of 0.25, confidence value of 0.5, and lift value of 0.5/0.25, which indicates an attraction; to be near the sun augments the probability of not having a moon. The two rules have identical supports and confidences but opposite lifts. In the sequel, we will use support as an indicator of the prevalence of a rule and lift as an indicator of its "truthness".

## 7.4. Data Mining for Fault Localization[6]

We consider a debugging process in which a program is tested against different test cases. Each test case yields a transaction in the AR sense, in which attributes correspond to properties observed during the execution of the test case. Two attributes, *PASS* and *FAIL*, represent the issue of the test case (again, see future works for variants on this). Thus, the set of all test cases yields a set of transactions that form a formal context, which we call a trace context. The main idea of the data mining approach is to look for a formal explanation of the failures.

### 7.4.1. Failure Rules

Formally, we are looking for association rules following pattern $P \rightarrow FAIL$. We call these rules *failure rules*. A failure rule proposes an explanation to a failure, and this explanation can be evaluated according to its support and lift. Note that failure rules have a variable premise $P$ and a constant conclusion $FAIL$. This slightly simplifies the management of rules. For instance, relevance indicators can be specialized as follows:

**Definition 5** (Relevance indicators for failure rules)

$$sup(P \rightarrow FAIL) = \|extent(P \cup \{FAIL\})\|,$$

$$conf(P \rightarrow FAIL) = \frac{\|extent(P \cup \{FAIL\})\|}{\|extent(P)\|},$$

$$lift(P \rightarrow FAIL) = \frac{\|extent(P \cup \{FAIL\})\| \times \|extent(\varnothing)\|}{\|extent(P)\| \times \|extent(\{FAIL\})\|}.$$

Observe that *extent(Ø)* and *extent({FAIL})* are constant for a given test suite. Only *extent(P)* and *extent (P ∪ {FAIL}* depend on the failure rule.

---
[6] Part of Chapter 7.4 is from Reference [6]

It is interesting to understand the dynamics of these indicators when new test cases are added to the trace context.

***Lemma 2*** (Dynamics of relevance indicators with respect to test suite). *Consider a failure rule P FAIL*:

*A new passed test case that executes P will leave its support unchanged (normalized support will decrease slightly[7]), will decrease its confidence, and will decrease its lift slightly if P is not executed by all test cases.*

*A new passed test case that does not execute P will leave its support and confidence unchanged (normalized support will decrease slightly) and will increase its lift.*

*A new failed test case that executes P will increase its support and confidence (normalized support will increase slightly) and will increase its lift slightly if P is not executed by all test cases.*

*A new failed test case that does not execute P will leave its support and confidence unchanged (normalized support will decrease slightly), and will decrease its lift.*

In summary, support and confidence grow with new failed test cases that execute $P$, and *lift* grows with failed test cases that execute $P$ or passed test cases that do not execute $P$. Failed test cases that execute $P$ increase all the indicators, but passed test cases that do not execute $P$ only increase *lift*[8].

Another interesting dynamic is what happens when $P$ increases.

***Lemma 3*** (Dynamics of relevance indicators with respect to premise). *Consider a failure rule P→FAIL and replacing P with P such that P>P*:

*Support will decrease (except if all test cases fail, which should not persist). One says P→FAIL is more specific than P→FAIL.*

Confidence and lift can go either way, but both in the same way because $\frac{extent(\emptyset)}{extent(\{FAIL\})}$ is a constant.

For the sequel of the description, we assume that the attributes recorded in the trace context are line numbers of executed statements. Since the order of the attributes in a formal context does not matter, this forms an abstraction of a standard trace (see a fragment of such a trace context in Table 7.4). Thus, explanations for failures will consist of line numbers, lines that increase the risk of failure when executed. Had other trace observations been used, the explanations would have been different. For faults that materialize in faulty instructions, it is expected that they will show up as explanations to failed test cases. For other faults that materialize in missing instructions, they will still be visible in actual lines that would have been correct if the missing lines where present. For instance, a missing initialization will be seen as the faulty consultation of a non-initialized variable. It is up to a competent debugger to conclude from faulty consultations that an initialization is missing. Note finally that the relationships

---

[7] Slightly: if most test cases pass, which they should eventually.

[8] Observing more white swans increases the belief that swans are white, but observing non-white non-swans increases the interest of the white swan observations. Observing a non- white swan does not change the support of the white swan observations, but it decreases its confidence and interest. Still, the interest can be great if there are more white swans and non-white non-swans than non-white swans.

between faults and failures are complex:

Table 7.4. A trace context [6].

| Test case | Executed lines | | | | Verdict | |
|---|---|---|---|---|---|---|
| | 57 | 58 | ... | 105 | PASS | FAIL |
| $t_1$ | × | × | | × | × | |
| $t_2$ | × | × | | × | | × |
| ... | ... | ... | ... | ... | ... | ... |

- Executing a faulty line does not necessarily cause a failure. For example, a fault in a line may not be stressed by a case test (e.g. faulty condition $i > 1$ instead of the expected $i > 0$, tested with $i$ equals to 10) or a faulty line that is "corrected" by another one.

- Absolutely correct lines can apparently cause failure, such as lines of the same basic block [29] as a faulty line (they will have exactly the same distribution as the faulty line) or lines whose preconditions cannot be established by a distant faulty part.

Failure rules are selected according to a minimal support criterion. However, there are too many such rules, and it would be inconvenient to list them all. We have observed in Lemma 3 that more specific rules have less support. However, this does not mean that less specific rules must be preferred. For instance, if the program has a mandatory initialization part, which always executes a set of lines $I$, rule $I \rightarrow FAIL$ is a failure rule with maximal support, but it is also less informative. On the contrary, if all failures are caused by executing a set of lines $F \supset I$, rule[9] $F\backslash I \rightarrow FAIL$ will have the same support as $F \rightarrow FAIL$, but will be the most informative. In summary, maximizing support is good, but it is not the definitive criteria for selecting informative rules.

Another idea is to use the lift indicator instead of support. However, lift does not grow monotonically with premise inclusion. Therefore, finding rules with a minimal lift cannot be done more efficiently than by enumerating all rules and then filtering them.

Table 7.5. Failure context for mutant 1 of the Trityp program with min $lift = 1.25$ and min $sup = 1$ (for mutant 1 the fault is at line 84, see Table 7.1 ) [6].

| Rule ID | Executed lines | | | | | | | | | | |
|---|---|---|---|---|---|---|---|---|---|---|---|
| | 17 | 58 | 66 | 81 | 84 | 87 | 90 | 105 | 93 | ... | 113 |
| $r_1$ | × | × | × | × | × | × | × | × | × | ... | × |
| $r_2$ | × | × | | × | × | × | × | | × | ... | × |
| ⋮ | ⋮ | ⋮ | ⋮ | ⋮ | ⋮ | ⋮ | ⋮ | ⋮ | ⋮ | ⋮ | ⋮ |
| $r_8$ | × | × | | × | | | | | × | ... | × |
| $r_9$ | × | × | | | | | | | × | ... | × |

### 7.4.2. Failure Lattice

Here, we describe how to use FCA to help navigate in the set of explanations.

**Definition 6** (The failure lattice). *Form a formal context with the premises of failure rules. The rules*

---

[9] Where .\. is set subtraction; the elements of a first set that do not belong to a second set.

*identifiers are the objects, and their premises are the attributes (in our example, line numbers) (see an example in* Table 7.5*). It is called the failure context.*

*Observe that the failure context is special in that all premises of failure rules are different from each other*[10]*. Thus, they are uniquely determined by their premises (or itemsets). Thus, it is not necessary to identify them by objects identifiers.*

*Apply FCA on this formal context to form the corresponding concept lattice. It is called the failure lattice. Its concepts and labeling display the most specific explanations to groups of failed tests.*

*Since object identifiers are useless, replace object labels by the support and lift of the unique rule that labels each concept. This forms the failure lattice (see* Figure 8.3*). The overall trace mining process is summarized in* Figure 8.4.

**Lemma 4** (Properties of the failure lattice). *The most specific explanations (i.e., the largest premises) are at the bottom of the lattice. On the contrary, the least specific failure rules are near the top. For instance, the line numbers of a prelude sequence executed by every test case will label the topmost concepts.*

*The explanations with the smallest support are at the bottom of the lattice. For example, line numbers executed only by specific failed test cases will label concepts near the bottom.*

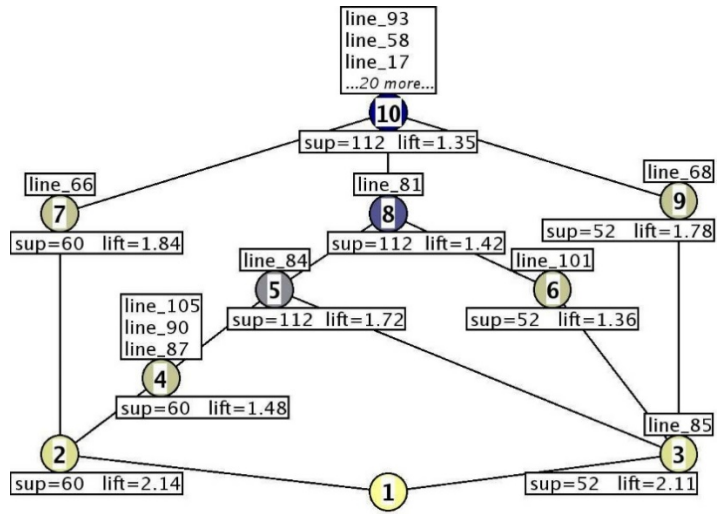

Figure 7.2. Failure lattice associated to the failure context of Table 7.5 (for mutant 1, the fault is at line 84) [6]

---

[10] This is a novel property with respect to standard FCA where nothing prevents two different objects to have the same attributes.

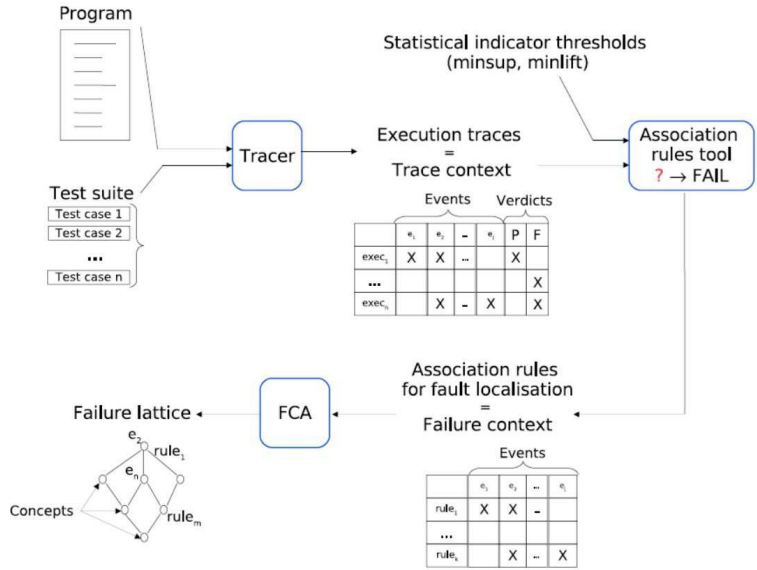

Figure 7.3. The trace mining process [6].

*Support increases when going upstream, from bottom to top. We call this the global monotony of support ordering. This is a theorem* [[5]].

*Lift does not follow any global monotony behavior.*

*Concepts form clusters of comparable concepts with the same support. For example, concepts 2, 4, and 7 in* Figure 7.2 *form a cluster of rules with support 60. We call them support clusters. This means that explanations of increasing size represent the same group of failures.*

*In a support cluster, a unique concept has the largest extent. We call it the head concept of the support cluster. It corresponds to the explanation with the highest lift value in the support cluster. More generally, lift decreases when going bottom-up in a support cluster. We call this behavior the local monotony of lift ordering, and it is also a theorem* [[5]].

*It is useless to investigate explanations other than the head concepts. This can be done by a bottom-up exploration of the failure lattice.*

In the lattice of Figure 7.2, only concepts 2 (head of support cluster with value 60), 3 (head of support cluster with value 52), and 5 (head of support cluster with value 112) need be presented to the debugging oracle. Concept 5 has Line 84 in its attribute label, which is the location of the fault in this mutant. The local monotony of lift ordering shows that the lift indicator can be used as a metric, but only inside support clusters.

The process that we have presented is dominated by the choice of a minimal value for the support indicator. Recall that the support of an explanation is simply the number of simultaneous realizations of its items in the failure context, and the normalized support is the ratio of this number to the total number of realizations. In this application of ARs, it is more meaningful to use the non-normalized variant because it directly represents the number of failed test cases covered by an explanation. What is a good value for the minimal support? First, it cannot be larger than the number of failed test cases (= *extent(FAIL)*); otherwise, no $P \rightarrow FAIL$ rule will show up. Second, it cannot be less than 1. The choice of between 1 and *extent(FAIL)* depends on the nature of the fault, but in any case, experiments show that acceptable minimum support is quite low (a few percent of the total number of test cases). A high minimal support will filter out all faults that are the causes of less failures than this threshold. Very singular faults will require a very small support, eventually 1, to be visible in the failure lattice. This suggests starting with a high support to localize the most visible faults, and then decreasing the support to localize less frequently executed faults. The minimal support acts as a resolution cursor. A coarse resolution will show the largest features at a low cost, and a finer resolution will be required to zoom in on smaller features at a higher cost.

We have insisted on using lift instead of confidence as a "truthness" indicator, because it lends itself more easily to an interpretation (recall Definition 4 and subsequent comments). However, in the case of failure rules, the conclusion is fixed (= *FAIL*), and both indicators increase and decrease in the same way when the premise changes (recall Lemma 3). The only difference is that the lift indicator yields a normalized value (1 is independence, bellow 1 is repulsion, and over 1 is attraction). What is the effect of a minimum lift value? Firstly, if it is chosen to be larger than or equal to 1, it will eliminate all failure rules that show a repulsion between the premise and conclusion. Secondly, if it is chosen to be strictly greater than 1, it will eliminate failure rules that have a lower lift, thus compressing the representation of support clusters and eventually eliminating some support clusters. Thus, the minimal lift also acts as a zoom.

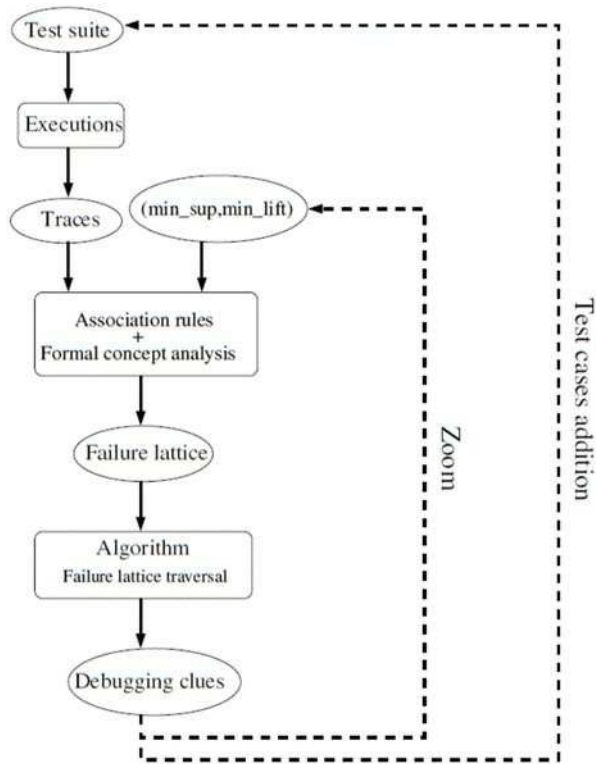

Figure 7.4. The global debugging process [6].

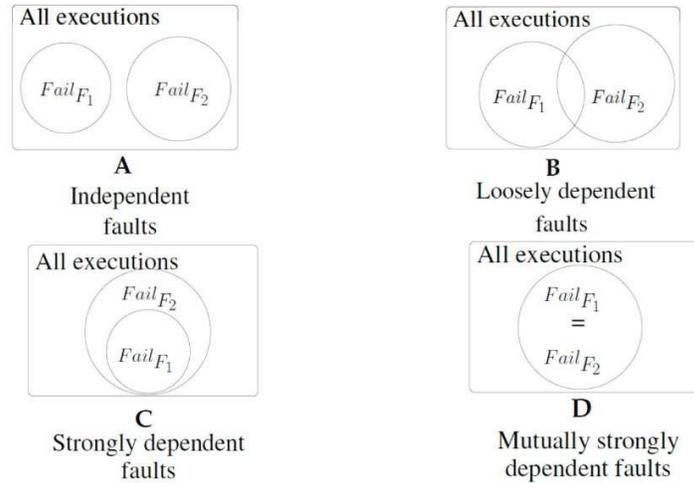

Figure 7.5. The four Venn diagrams of two-fault dependency [6].

This suggests a global debugging process in which the results of an increasingly large test suite are examined with increasing acuity (see
Figure 7.4). Given a test suite, an inner loop computes failure rules, i.e., explanations with decreasing support, from a fraction of *extent(FAIL)* to 1, and builds the corresponding failure lattice. In the outer loop, test cases are added progressively to cope with added functionality (such as test-driven development) or new failure reports. Thus, the global debugging process zooms in on the failed test cases to find explanations for more and more specific failures.

## 7.5. The Failure Lattice for Multiple Faults[11]

This section extends the analysis of data mining for fault localization for the multiple fault situation. From the debugging process point of view, there is nothing special about multiple faults. Some software engineering life cycles like test-driven development tend to limit the number of faults observed simultaneously, but one can never assume a priori that there is a single fault. Thus, we assume there are one or several faults.

### 7.5.1. Dependencies between Faults

In the multiple fault case, each failure trace accounts for one or several faults. Conversely, faulty lines are suspected in one or several failure traces. Thus, the inner loop of the global debugging process cannot stop because a fault is found. The process must go on until all failures are explained. How can this be done without exploring the entire failure lattice?

Consider any pair of two faults $F_1$ and $F_2$, and $Fail_{F_1}$ and $Fail_{F_2}$ are the sets of failed test cases that detect $F_1$ and $F_2$, respectively. We identify four types of possible dependencies between the two faults.

Definition 7 (Dependencies between faults) If $Fail_{F_1} = Fail_{F_2}$, we say that they are mutually strongly dependent (MSD).

If $Fail_{F_1} \not\subset Fail_{F_2}$ we say $F_1$ is strongly dependent (SD) from F2 (and vice versa).

---

[11] Part of Chapter 7.5 is from Reference [6]

If $Fail_{F_1} \cap Fail_{F_2} \neq \emptyset$, we say that they are loosely dependent (*LD*).

Otherwise, $Fail_{F_1} \cap Fail_{F_2} = \emptyset$, we say that they are independent (*ID*).

Note that this classification, is not intrinsic to a pair of faults; it depends on the test suite. However, it does not depend arbitrarily from the test suite.

Lemma 5 (How failure dependencies depend on growing test suites) Assume that the test suite can only grow, then an *ID* or *SD* pair can only become *LD*, and an *MSD* pair can only become *SD* or *LD*.

This can be summarized as follows:

$$ID \rightarrow LD \leftarrow SD \leftarrow MSD$$

Note also that this knowledge, with several faults and the dependencies between them, is what the debugging person is looking for, whereas the trace context only gives hints at this knowledge. The question is: how does it give hints at this knowledge?

The main idea is to distinguish special concepts in the failure lattice that we call *failure concepts*.

*Definition 8* (Failure concept). *A failure concept is a maximally specific concept of the failure lattice whose intent (a set of lines) is contained in a failed execution.*

Recall that the failure rules are an abstraction of the failed execution. For instance, choosing minimal support and lift values eliminates lines that are seldom executed or that do not attract failure. Thus, the failure lattice describes exactly the selected failure rules but only approximately the failed executions. That is why it is interesting; it compresses information, though with loss. The failure concepts in the failure lattice are the concepts that best approximate failed executions. All other concepts contain less precise information. For the same reasons, there are much fewer failure concepts than failed executions; each failure concept accounts for a group of failures that detects some fault.

The main use for failure concepts is to give a criterion for stopping the exploration of the failure lattice. In a few words,

- The bottom-up exploration of the failure lattice goes from support clusters to support clusters as above;

- The line labels of the traversed concepts are accumulated in a fault context sent to the competent debugger;

- Any time a competent debugger finds a hint at an actual fault, all the failure concepts under the concept that gave the hint are deemed explained;

- The process continues until all failure concepts are explained.

The fault context is the part of the program that the debugging person is supposed to check. We consider its size as a measure of the effort imposed on the debugging person (see Section 1.6 for comparative experiments).

Dependencies between faults have an impact on the way failure concepts are presented in the failure lattice.

***Lemma 6*** (*ID* faults with respect to failure concepts). *If two faults are ID, their lines can never occur in the same failed trace, and then no rule contains the two faults and no concept in the failure lattice contains the two faults. Thus, the two faults will label failure concepts in two different support clusters that have no subconcepts in common (for an example, see* Figure 7.6*).*

Concretely, when exploring the failure lattice bottom-up, finding a fault in the label of a concept explains both the concept and the concepts underneath, but the faults in the other upper branches remain to be explained. Moreover, the order with which the different branches are explored does not matter.

***Lemma 7*** (*LD* faults with respect to failure concepts). *If two faults are LD, some failed traces contain both faults, while other failed traces contain either fault. They may label concepts in two different support clusters that share common subconcepts.*

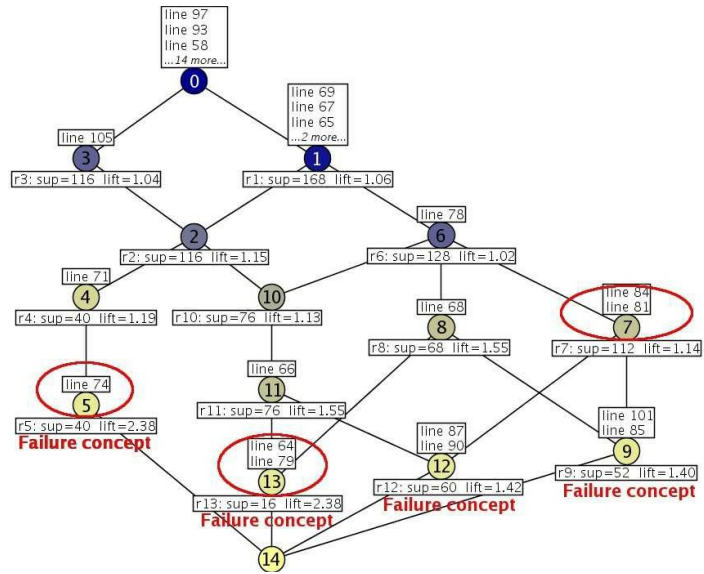

Figure 7.6. Failure lattice associated to program Trityp with ID faults of mutants 1, 2, and 6 [6].

Concretely, when exploring the failure lattice bottom-up, finding a fault for a failure concept does not explain the other *LD* failure concept. Once a fault is found, shared concepts must be re-explored in the direction of other superconcepts.

***Lemma 8*** (*SD* faults with respect to failure concepts). *If two faults are SD, say F1 depends on F2, a failure concept whose intent contains LineF1 will appear as a subconcept of a failure concept whose concept contains LineF2 in a different support cluster (for an example, see* Figure 7.7*).*

Therefore, fault $F_1$ will be found before $F_2$, but the debugging process must continue because there is a failure concept above.

***Lemma 9*** (*MSD* faults with respect to failure concepts). *Finally, if two faults are MSD, they cannot be distinguished by failed executions, and their failure concepts belong to the same support cluster. However, they can sometimes be distinguished by passed executions (such as one having more passed executions than the other), and this can be seen in the failure lattice through the lift value.*

All this can be formalized in an algorithm that searches for multiple faults in an efficient traversal of the failure lattice (see Algorithm 7.1).

```
1  C_toExpolore := FAILURECONCEPTS
2  C_failure_toExplain := FAILURECONCEPTS
3  while C_failuretoExplain ≠ ∅ ∧ C_toExplore ≠ ∅ do
4      let c ∈ C_toExplore in
5      C_toExplore := C_toExplore \ {c}
6      if the debugging oracle(label(c), faultcontext(c)) locates no fault
         then
7        |  C_toExplore := C_toExplore ∪ {upperneighboursofc}
8      else
9         let Explained = subconcepts(c) ∪ cluster(c) in
10        C_toExplore := C_toExplore \ Explained
11        C_failuretoExplain := C_failuretoExplain \ Explained
12     end
13 end
```

Algorithm 7.1). The failure lattice is traversed bottom-up, starting with the failure concepts (step 1). At the end of the failure lattice traversal, $C_{failuretoExplain}$, the set of failure concepts not explained by a fault (step 2) must be empty, or all concepts must be already explored (step 3). When a concept, $c$ (step 4), is chosen among the concepts to explore, $C_{toExplore}$, the events that label the concept are explored. Note that the selection of that concept is not determinist. If no fault is located, then the upper neighbours of $c$ are added to the set of concepts to explore (step 7). If, thanks to the new clues, the debugging oracle understands mistakes and locates one or several faults, then all subconcepts of c and all concepts that are in the same support cluster are "explained". Those concepts do not have to be explored again (step 10). This means that the failure concepts that are subconcepts of $c$ are explained (step 11). The exploration goes on until all failed executions in the failure lattice are explained by at least one fault, or all concepts have been explored.

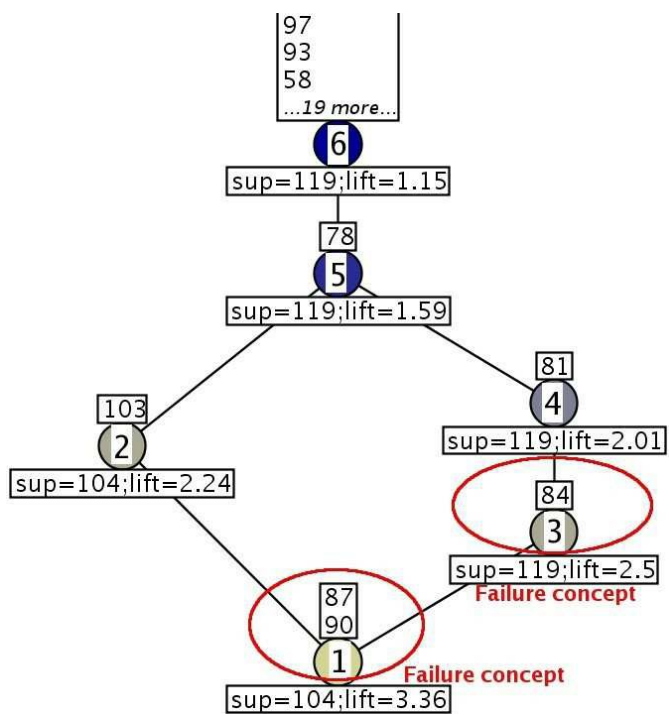

Figure 7.7. Failure lattice associated to program Trityp with SD faults 1 and 7 [6].

```
1  C_toExpolore := FAILURECONCEPTS
2  C_failure_toExplain := FAILURECONCEPTS
3  while C_failuretoExplain≠∅ ∧ C_toExplore≠∅ do
4  |   let c ∈ C_toExplore in
5  |   C_toExplore := C_toExplore\{c}
6  |   if the debugging oracle(label(c), faultcontext(c)) locates no fault
       then
7  |   |   C_toExplore := C_toExplore ∪ {upperneighboursof c}
8  |   else
9  |   |   let Explained = subconcepts(c) ∪ cluster(c) in
10 |   |   C_toExplore := C_toExplore\Explained
11 |   |   C_failuretoExplain := C_failuretoExplain\Explained
12 |   end
13 end
```

Algorithm 7.1. Failure lattice traversal [6].

Table 7.6. Exploration of the failure lattice of Figure 7.6 [6].

| Iteration | $C_{toExplore}$ | $C_{failuretoExplain}$ |
|---|---|---|
| 0 | $\{c_5, c_{13}, c_{12}, c_9\}$ | $\{c_5, c_{13}, c_{12}, c_9\}$ |
| 1 | $\{c_{13}, c_{12}, c_9\}$ | $\{c_{13}, c_{12}, c_9\}$ |
| 2 | $\{c_{12}, c_9\}$ | $\{c_{12}, c_9\}$ |
| 3 | $\{c_9, c_7, c_{11}\}$ | $\{c_{12}, c_9\}$ |
| 4 | $\{c_7, c_{11}, c_8\}$ | $\{c_{12}, c_9\}$ |
| 5 | $\{c_{11}, c_8\}$ | {} |

Note that at each iteration, $C_{failuretoExplain}$ can only decrease or remain untouched. The competent debugger hypothesis ensures that $C_{failuretoExplain}$ ends at empty when min sup is equal to 1. In case of an incompetent debugging oracle or a too high min sup, the process would end when $C_{toExplore}$ becomes empty, namely when all concepts have been explored.

### 7.5.2. Example

For the example of Figure 7.6, the min sup value is equal to four failed executions (out of 400 executions, of which there are 168 failed executions), and the min lift value is equal to one. There are four failure concepts: 5, 13, 12, and 9. Table 7.6 presents the values of $C_{toExplore}$ and $C_{failuretoExplain}$ at each iteration of the exploration. We choose to explore the lattice with a queue strategy; it means first in $C_{toExplore}$, first out of $C_{toExplore}$. However, the algorithm does not specify one strategy.

At the begining, $C_{toExplore}$ and $C_{failuretoExplain}$ are initialized as the set of all failure concepts (iteration 0 in Table 7.6). At the first iteration of the while loop, concept 5 is selected ($c = c_5$). That concept is labeled by line 74. Line 74 actually corresponds to fault 6. Thanks to the competent debugging hypothesis, fault 6 is located. Concepts 5, 4, and 14 are thus tagged as explained. The new values of $C_{toExplore}$ and $C_{failuretoExplain}$ are presented at iteration 1 in Table 7.6.

At the second iteration, concept 13 is selected ($c = c_{13}$). That concept is labeled by lines 64 and 79. Line 79 actually corresponds to fault 2; the competent debugging oracle locates fault 2. Concept 13 is tagged as explained. At the third iteration, concept 12 is selected. That concept is labeled by lines 87 and 90. No fault is found. The upper neighbours, concepts 7 and 11, are added to $C_{toExplore}$, and $C_{failuretoExplain}$ is unchanged.

At the next iteration, concept 9 is selected. As in the previous iteration, no fault is found. The upper neighbour, concept 8, is added to $C_{toExplore}$.

Finally, concept 7 is selected. That concept is labeled by lines 81 and 84. By exploring those lines (new clues) in addition with the fault context, i.e., lines that have already been explored: 87, 90, 101, and 85, the competent debugging oracle locates fault 1 at line 84. The fault is the substitution of the test of trityp = 2 by trityp = 3. Concepts 12 and 9 exhibit two concrete realizations (failures) of the fault at line 84 (Concept 7). Concepts 7, 12, and 9 are tagged as explained. The set of failure concepts to explain is empty; thus, the exploration stops. All four faults (for failures above support and lift threshold) are found after the debugging oracle has inspected nine lines.

## 7.6. Discussion[12]

The contexts and lattices introduced in the previous sections allow programmers to see all the differences between execution traces as well as all the differences between association rules. There exist other methods that compute differences between execution traces. We first show that the information about trace differences provided by the failure context (and the corresponding lattice) is already more relevant than the information provided by four other methods proposed by Renieris and Reiss [25] and Cleve and Zeller [7]. Then, we show that explicitly using association rules with several lines in the premise alleviates some limitations of Jones et al.'s method [15]. Finally, we show that reasoning on the partial ordering given by the proposed failure lattice is more relevant than reasoning on total order rankings [8][18][20][24][32].

### 7.6.1. The Structure of the Execution Traces

The trace context contains the whole information about execution traces. In particular, the associated lattice, the trace lattice, allows programmers to see all differences between traces in one pass.

There exist several fault localization methods based on the differences between execution traces. They all assume a single failed execution and several passed executions. We rephrase them in terms of search in a lattice to highlight their advantages, their hidden hypothesis, and limitations.

### 7.6.2. Union Model

The union model, proposed by Renieris and Reiss [25], aims at finding features that are specific to the failed execution. The method is based on trace differences between the failed execution f and a set of passed executions $S : f - \cup_{s \in S} s$. The underlying intuition is that the failure is caused by lines that are executed only in the failed execution. Formalized in FCA terms, the concepts of interest are the subconcepts whose label contains *FAIL*, and the computed information is the lines contained in the labels of the subconcepts. The trace lattice presented in the figure is slightly different from the lattice that would be computed for the union model, because it represents more than one failed execution. Nevertheless, the union model often computes empty information, namely each time the faulty line belongs to failed and passed execution traces. For example, a fault in a condition has a very slight chance to be localized. The approach we presented is based on the same intuition. However, the lattices that we propose do not lose information and help navigate in order to localize the faults, even when the faulty line belongs to both failed and passed execution traces.

---

[12] Part of Chapter 7.6 is from Reference [6]

The union model helps localize a bug when executing the faulty statement always implies an error, such as the bad assignment of a variable that is the result of the program. In that case, the lattice also helps, and the faulty statement labels the same concept as *FAIL*.

### 7.6.3. Intersection Model

The intersection model [25] is the complementary of the previous model. It computes the features whose absence is discriminant of the failed execution: $\bigcap_{s \in S} s - f$. Replacing *FAIL* with *PASS* in the above discussion is relevant to discussing the intersection model and leads to the same conclusions.

### 7.6.4. Nearest Neighbor

The nearest neighbor approach [25] computes a distance metric between the failed execution trace and a set of passed execution traces. The computed trace difference involves the failed execution trace, $f$, and only one passed execution trace, the nearest one, $p : f - p$. The difference is meant to be the part of the code to explore. The approach can be formalized in FCA. Given a concept $C_f$ whose intent contains *FAIL*, the nearest neighbor method searches for a concept $C_p$ whose intent contains *PASS*, such that the intent of $C_p$ shares as many lines as possible with the intent of $C_f$.

The rightmost concept fails, whereas the leftmost one passes. As for the previous methods, it is a good approach when the execution of the faulty statement always involves an error. However, when the faulty statement can lead to both a passed and a failed execution, the nearest neighbor method is not sufficient. In addition, we remark that there are possibly many concepts of interest, namely all the nearest neighbors of the concept that is labeled by *FAIL*. With a lattice, that kind of behavior can be observed directly.

Note that in the trace lattice, the executions that execute the same lines are clustered in the label of a single concept. Executions that are nearby share a large part of their executed lines and label concepts that are neighbors in the lattice. There is therefore no reason to restrict the comparison to a single pass execution. Furthermore, all the nearest neighbors are naturally in the lattice.

### 7.6.5. Delta Debugging

*Delta debugging*, proposed by Zeller et al. [8], reasons on the values of variables during executions rather than on executed lines. The trace spectrum, and therefore the trace context, contains different types of attributes. Note that the presented approach does not depend on the type of attributes and would apply more to spectra containing other attributes than executed lines.

Delta debugging computes the differences between the failed execution trace and a single passed execution trace in a memory graph. By injecting the values of variables of the failed execution into variables of the passed execution, the method tries to determine a small set of suspicious variables. One of the purposes of the method is to find a passed execution relatively similar to the failed execution. It has the same drawbacks as the nearest neighbor method.

### 7.6.6. From the Trace Context to the Failure Context

Tarantula
Jones et al. [15] computed association rules with only one line in the premise. Denmat et al. [10] showed the limitations of this method through three implicit hypotheses. The first hypothesis is that a failure has a single faulty statement origin. The second hypothesis is that lines are independent. The third hypothesis is that executing the faulty statement often causes a failure. That last hypothesis is a common assumption of fault localization methods, including the presented method. Indeed, when the fault is executed in both passed and failed executions (such as in a prelude), it cannot be found so easily using these hypotheses.

In addition, Denmat et al. demonstrated that the ad hoc indicator, which was used by Jones et al., is equivalent to the lift indicator.

By using association rules with more expressive premises than in Jones et al.'s method (namely with several lines), the limitations mentioned above are alleviated. Firstly, the fault need not be a single line but can contain several lines together. Secondly, the dependency between lines is taken into account. Indeed, dependent lines are clustered or ordered together.

The part of the trace context that is important to search in order to localize a fault is the set of concepts that are related to the concept labeled by *FAIL*; i.e., those that have a non-empty intersection with the concept labeled by *FAIL*. Computing association rules with *FAIL* as a conclusion compute exactly those concepts, modulo the *minsup* and *minlift* filtering. In other words, the focus is done on the part of the lattice related to the concept labeled by *FAIL*.

### 7.6.7. The Structure of Association Rules

Jones et al.'s method presents the result of the analysis to the user as a coloring of the source code. A red-green gradient indicates the correlation with failure. Lines that are highly correlated with failure are colored in red, whereas lines that are not highly correlated are colored in green. Red lines typically represent more than 10% of the lines of the program, without identified links between them. Other statistical methods [8][18][19][32] also try to rank lines in a total ordering. It can be seen as ordering the concepts of the failure lattice by the lift value of the rule in their label. However, we have shown in Section 1.3 that the monotonicity of lift is only relevant locally to a support cluster.

For example, on the failure lattice of Figure 7.2, the obtained ranking would be: line 85, line 66, line 68, line 84. No link would be established between the execution of line 85 and line 68, for example.

The user who must localize a fault in a program has background knowledge about the program and can use it to explore the failure lattice. Reading the lattice gives context about the fault and not just a sequence of independent lines to be examined, and it reduces the number of lines to be examined at each step (concept) by structuring them.

### 7.6.8. Multiple Faults

We have compared the failure lattice with existing single fault localization methods. In this section, we compare the presented navigation in the failure lattice with the strategies of the other methods to detect several faults.

The presented approach involves algorithmic debugging [27]. The difference lies in the traversed data structure. While Shapiro's algorithm helps traverse a proof tree, the presented algorithm helps traverse the failure lattice, starting from the most suspicious places.

For multiple faults, Jiang et al. [14] criticized the ranking of statistical methods. They proposed a method based on traces whose events are predicates. The predicates are clustered, and the path in the control flow graph associated to each cluster is computed. In the failure lattice, events are also clustered in concepts. The relations between concepts give information about the path in the control flow graph and highlight some parts of that path as relevant to debug without computing the control flow graph.

Zheng et al. [32] proposed a method based on bi-clustering in order to group failed executions and identify one feature (bug predictor) that characterizes each cluster. They proposed to look at one bug predictor at a time. Several bug predictors can be in relation with the same fault, but no link is drawn between them. The presented approach gives context to the fault, in order to help understand the mistakes of the programmer that have produced the fault.

Jones et al. [16] proposed a method that first clusters executions and then finds a fault in each cluster in parallel. The method has the same aim as the presented method. Indeed, in both cases, we want to separate the effects of the different faults in order to treat the maximum of faults in one execution of the test suite, but in the presented approach, the clusters are partially ordered to take into account dependencies between faults.

Finally, SBI [18] introduces a stop criterion as we did in the presented algorithm. SBI tries to take advantage of one execution of the test suite. The events are predicates, and SBI ranks them. When a fault is found due to the ranking, all execution traces that contain the predicates used to find the fault are deleted, and a new ranking of predicates with the reduced set of execution traces is computed. Deleting execution traces can be seen as equivalent to tagging concepts, as well as the events of their labeling, as explained in DeLLIS. The difference between SBI and DeLLIS is that DeLLIS does not need to compute the failure lattice several times.

## 7.7. Fault Localization using N-gram Analysis[13]

In the previous sections, we described the background of data mining and how it can be applied to fault localization in general. In this section, we will describe how to use data mining along with N-gram analysis for software fault localization.

In software fault localization, test cases are usually utilized as sets of inputs with known expected outputs. If the actual output does not match the expected output, the test case has failed. Various information can be collected during the execution of the test cases for later analysis. This information may include statement coverage (the set of statements that were executed at least once during the execution) and exact execution sequence (the actual order in which the statements were executed during the test case executions). Since we will be working only with the exact execution sequence in this paper, we refer to it as trace. Usually, the usefulness of trace data is limited by the sheer volume. Data mining traditionally deals with large volumes of data, and in this research, we apply data mining techniques to process this trace data for fault localization. From trace data, we generate N-grams, i.e., subsequences of length N. From these, we choose N-grams that appear more than a certain number of times in the failing traces. For these N-grams, we calculate the confidence — the conditional probability that a test case fails given that the N-gram appears in that test case's trace. We sort the N-grams in descending order of confidence and report the statements in the program in the order of their first occurrence in the sorted list.

### 7.7.1. Background

Execution Sequence
Let $P$ be a program with n lines of source code, labeled $L = \{l_1, l_2, \cdots, l_n\}$. For example, in the sample program mid from [15] in Figure 7.8, L= {4,5,6,10,11,12,13,14,15,17,18,19,20,21,24} after excluding comments, blank lines and structural constructs like '}'. A test case is a set of input with known outputs. Let $T = \{t_1, t_2, \cdots, t_n\}$ be the n test cases for program P. Each test case $t_i = \langle I_i, X_i \rangle$ has the input $I_i$ and expected output $X_i$. When program $P$ is executed with input $I_i$, it produces actual output $A_i$. If $A_i = X_i$, then we say $t_i$ is a passing test case, and if $A_i \neq X_i$; then we say $t_i$ is a failing test case. For example, the 6 test cases for the program mid in [15], $T = (t_1, t_2, \cdots, t_6)$, are shown in Table 7.7. Let $Y = \langle y_1, y_2, \cdots, y_k \rangle$ be the trace of program $P$ when running test case $T$. Then, for mid the trace for the test case $t_1$ is Y = {4,4,5,10,11,12,14,15,24,6}. We define two sets based on the outcome of the test cases- passing traces which is $Y_P = \{Y_i | t_i\, sa\, failing\, testcase\}$ and failing traces which is $Y_F = \{Y_i | t_i\, sa\, failing\, testcase\}$.

---
[13] Part of Chapter 7.7 is from Reference [22]

We define the problem as: *given program P with executable statements L ,test cases T and actual outputs A , the problem is to rank the statements in L according to their probability of containing the fault.* To compare this method with other methods like [15], we report the results in terms of statements, but it can also work at function level.

Given an ordered list, an N-gram is any sub-list of N consecutive elements in the list. The elements of the N-gram must be in the same order as they were in the original list, and they must be consecutive. Given an execution trace $Y$, an N-gram $G_{Y,N,\alpha}$ is a contiguous subsequence $\langle y_\alpha, y_{\alpha+1}, \cdots, y_{\alpha+N-1} \rangle$ of length N starting at position $\alpha$. For a trace $Y$, the set of all line N-gram is
$G_{Y,N} = \{G_{Y,N,1}, G_{Y,N,2}, \cdots, G_{Y,N,K-N+1}\}$

Table 7.7. Test cases for program *mid* [22]

| $t_i$ | $I_i$ | $X_i$ | $A_i$ | Results | trace |
|---|---|---|---|---|---|
| $t_1$ | 3,3,5 | 3 | 3 | Pass | 4,4,5,10,11,12,14,15,24,6 |
| $t_2$ | 1,2,3 | 2 | 2 | Pass | 4,4,5,10,11,12,13,24,6 |
| $t_3$ | 3,2,1 | 2 | 2 | Pass | 4,4,5,10,11,18,13,24,6 |
| $t_4$ | 5,5,5 | 5 | 5 | Pass | 4,4,5,10,11,18,13,24,6 |
| $t_5$ | 5,3,4, | 4 | 4 | Pass | 4,4,5,10,11,12,13,24,6 |
| $t_6$ | 2,1,3 | 2 | 1 | Fail | 4,4,5,10,11,12,14,15,24,6 |

### 7.7.2. Linear Execution Blocks

From the set of all traces, we identify the execution blocks, i.e., the code segments with a single point of entry and a single point of exit. For this, we construct the Execution Sequence Graph XSG(P) = (V, E) where the set of vertices is $V \subseteq L$ such that for each $v_i \in V$, $v_i \in Y_k$ for some $k$ and that $v_i$ and $v_j$ are consecutive in $Y_k$. This is similar to a Control Flow Graph, but the vertices in an XSG represent statements rather than blocks. In this graph, there is an edge between two vertices only if they were executed in succession in at least one of the execution traces. The XSG for mid is given in Figure 8.10, where we can see that the blocks of mid are $\{b_1, b_2, \cdots, b_{10}\} = \{\langle 4 \rangle, \langle 5,10,11 \rangle, \langle 12 \rangle, \langle 18 \rangle, \langle 20 \rangle, \langle 24,6 \rangle, \langle 14 \rangle, \langle 13 \rangle, \langle 15 \rangle\}$. Thus, trace of test case t1 can be converted to block level trace by $\langle b_1, b_2, b_3, b_8, b_{10}, b_7 \rangle$.

It should be noted that the definition of blocks here is different than the traditional blocks [[1]]. Since we identify blocks from traces, the blocks here may include function or procedure entry points. For example, $\langle 5,10,11 \rangle$ will not be a single block by the traditional definition since it has a function started at line 10. Due to this difference, we name the blocks Linear Execution Blocks, defined as follows: A Linear Execution Block $B\langle v_i, v_{i+1}, \cdots, v_j \rangle$ is *a directed path in XSG such that the indegree of each vertex*

$V_k \in B$ is 0 or 1. Advantages of using block traces are: (a) it reduces the size of the traces, and, (b) in a block trace, each sequence of two blocks indicate one possible branch. Therefore, in N-gram analysis on block traces, each block N-gram represents N - 1 branches. This helps the choice of N for N-gram analysis.

```
1   #include <stdio.h>
2   int main(){
3       int x, y, z, m;
4       scanf("%d %d %d, ",&x, &y, &z);
5       m = mid(x, y, z);
6       printf("%d",m);
7   }
8   int mid(int x, int y, int z){
9       int m;
10      m = z;
11      if (y<z){
12          if (x<y){
13              m = y;
14          }else if (x<z){
15              m = y;
16          }
17      }else{
18          if (x>y){
19              m = y;
20          }else if (x>z){
21              m = x;
22          }
23      }
24      return m;
25  }
```

Figure 7.8. Sample source code: mid.c [22]

### 7.7.3. Association Rule Mining

*Association rule mining* searches for interesting relationships among items in a given data set [13]. It has the following two parts:

Frequent Itemset Generation. Search for sets of items occurring together frequently, called a *frequent itemset*, whose frequency in the data set, called *support*, exceeds a predefined threshold, called *minimum support*.

Association Rule Generation. Look for association rules like $A \rightarrow B$ among the elements of the frequent itemsets, meaning that the appearance of $A$ in a set implies the appearance of $B$ in the same set. The conditional probability $P(B|A)$ is called *confidence*, which must be greater than a predefined *minimum confidence* for a rule to be considered. More details can be found in [13].

We model the blocks as items and the block traces as transactions. For example, $Y_1 = \langle b_1, b_2, b_3, b_8, b_{10}, b_7 \rangle$ is a transaction for *mid* corresponding to the first test case, $T_1$. We generate frequent itemsets from the transactions with the additional constraint that the items in an itemset must be consecutive in the original transaction. To do this, we generate $N$-grams from the block traces, and from them, we choose the ones with at least the minimum support. For a block $N$-gram $G_{Y_i,N,p}$, support is the number of failing traces containing $G_{Y_i,N,p}$:

$$Support(G_{Y_i,N,p}) = \left| \{Y_j | G_{Y_i,N,p} \in Y_j \text{ and } Y_j \in Y_F \} \right|$$

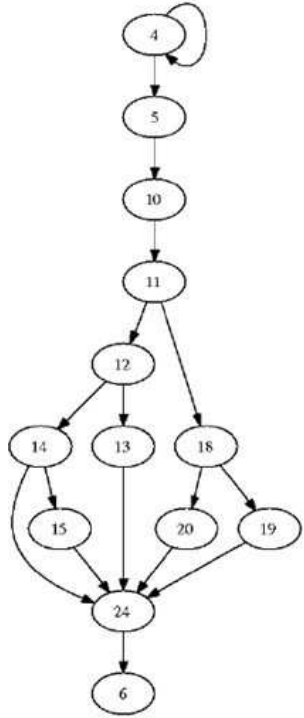

Figure 7.9. Execution sequence graph for program mid [22].

For example, for mid, the support for $\langle b_2, b_3, b_8 \rangle$ is 1 since it occurs in one failing trace. We add the test case type to the itemset. For example, after adding the test case type to the itemset $\langle b_2, b_3, b_8 \rangle$, the itemset becomes $\langle b_2, b_3, b_8, passing \rangle$. Then, we try to discover association rules of the form $A \Rightarrow failing$ from these itemsets where the antecedent is a block N-gram and the consequent is failing. Therefore, the block N-grams that appear as antecedents in the association rules are most likely to have caused the failure of the test case. We sort these block N-grams in descending order of confidence. For a block N-gram $G_{Y_i,N,p}$; confidence is the conditional probability that the test case outcome is failure given that $G_{Y_i,N,p}$ appears in the trace of that test case. That is,

$$Confidence\left(G_{Y_i,N,p}\right) = \frac{\Prob\left(G_{Y_i,N,p} \in Y_j \text{ and } t_j \text{ is a failing test case}\right)}{\Prob\left(G_{Y_i,N,p} \in Y_j\right)}$$

For example, the confidence the rule $\langle b_2, b_3, b_8 \rangle \Rightarrow failing$ has confidence 0.33. After sorting the block N-grams, we convert the blocks back to line numbers and report this sequence of lines to investigate to find the fault location.

### 7.7.4. Methodology

As input, we use the source code, the test case types, and the traces for all the test cases, and we produce as output an ordered list of statements, sorted in order of probability of containing the fault. We first convert the traces to block traces and then apply *N*-gram analysis on these block traces to generate all possible unique *N*-grams for a given range of *N*. For each *N*-gram, we count its frequency in passing and failing traces.

The execution of the faulty statement may not always cause failure of the test case. There may be quite a number of test cases in which the faulty statement was executed but did not cause a failure. In most cases, the failure is dependent on the sequence of execution. A specific sequence or path of execution will cause the program to fail, and this sequence will be very common in the failing traces but not so common in the passing traces. Therefore, we can find the subsequences that are most likely to contain the fault by analyzing the traces during passing and failing test cases.

There are two major parameters in the algorithm. The first one is *MinSup*, the minimum support for selecting the *N*-grams, and the second is $N_{MAX}$, the maximum value of *N* for generating the *N*-grams. Taking a low value of mini-mum support will result in the inclusion of irrelevant *N*-grams in consideration. Therefore, we should take minimum support at a high value. Our experience suggests that 90% is a good choice. However, the choice of an appropriate $N_{MAX}$ is more difficult. Two execution paths can differ because of conditional branches. Such differences can be detected by 2-grams. Again, the same function can be called from different functions, which can also be detected with 2-grams. Since we are using execution blocks, an *N*-gram can capture (*N* 1) branches, and a choice of 2 or 3 for $N_{MAX}$ should give good results in most cases. If we use higher *N*-grams, the algorithm will still be able to find the fault, but due to larger *N*-grams, we will have to examine more lines to find the fault.

L2B: Convert exact execution sequences to block traces. From the line level traces, we create the execution sequence graph (*XSG*). From the *XSG*, we find the linear execution blocks (*LEB*). Then, we convert the traces into block traces in lines 2 to 4 of Algorithm 7.2.

GNG: Generate *N*-grams. In this step, we first generate all possible *N*-grams of lengths 1 to $N_{MAX}$ from the block traces. The generation of all *N*-grams from a set of block traces for a given *N* is done in lines 1 to 7, and the generation and combination of all the *N*-grams are done in lines 5 to 8. Then, we find out how many passing and failing traces each *N*-gram occurs in.

FRB: Find relevant blocks. From 1-gram, we construct a set of relevant blocks, $B_{rel}$, that contains only the blocks that have appeared in each of the failing traces in lines 10 to 14.

EIN: Eliminate irrelevant *N*-grams. In lines 15 to 16, we discard the

*N*-grams that do not contain any block from the relevant block set, $B_{rel}$.

FFN: Find frequent *N*-grams. In lines 17 to 21, we eliminate *N*-grams with support less than the minimum support.

RNC: Rank *N*-grams by confidence. For each surviving *N*-gram, we compute its confidence using Equation (2). This is done in lines 22 to 26. Then, we order the *N*-grams in order of confidence in line 27.

B2L: Convert blocks in N-grams to line numbers. We convert each block in the *N*-grams back to line numbers using the *XSG* in line 28.

RLS: Rank lines according to suspicion. We traverse the ordered list of *N*-grams and report the line numbers in the order of their first appearance in the list. This is done in line 29.

If there are multiple *N*-grams with the same confidence as the *N*-gram containing the faulty statement,

the best case will be the ordering in which the faulty statement appears in the earliest possible position in the group, and the worst case will be the ordering in which the faulty statement appears in the latest possible position.

```
 1  Function LocalizeFaults(Y, Y_F, K, MINSUP):
 2    foreach Y_i ∈ Y do
 3      Convert Y_i to block trace
 4    end
 5    NG ← φ for N = 1 to N_MAX do
 6      NG ← NG ∪ GenerateNGrams(Y, N)
 7    end
 8    L_rel ← {n|n ∈ NG and |n| = 1} foreach n ∈ L_rel do
 9      if Support(n) ≠ |Y_F| then
10        Remove n from NG and L_rel
11      end
12    end
13    NG_1 ← {n|n ∈ NG and for all s ∈ L_rel, s ∉ n}
14    NG ← NG − NG_1
15    foreach n ∈ NG do
16      if Support(n) < MINSUP then
17        Remove n from NG
18      end
19    end
20    foreach n ∈ NG do
21      NF ← |{Y_k|Y_k ∈ Y_F and n ∈ Y_k}|
22      NT ← |{Y_k|Y_k ∈ Y and n ∈ Y_k}|
23      n.confidence ← NF ÷ NT
24    end
25    Sort NG in the descending order of confidence
26    Convert the block numbers in the N-grams in NG to line numbers
27    Report the line numbers in the order of their first appearance in NG
```

Algorithm 7.2. Fault localization using *N*-gram analysis [22].

```
 1  Function GenerateNGrams(Y, N):
 2    G ← φ
 3    foreach Y_i ∈ Y do
 4      G ← G ∪ G_{Y_i, N}
 5    end
 6    return G
```

Algorithm 7.3. N-gram generation [22].

### 7.7.5. Conclusion

In this section, we presented an approach to locate the suspiciousness statement using *N*-gram analysis. The augmenting the execution traces with data flows in order to pinpoint data-driven faults is worth investigating. Different than other fault localization techniques, such as spectrum, the amount of data that *N*-gram requires is relatively small; this gives it an advantage in the fault localization world, where the complexity and size of software are growing dramatically. Research using exact execution sequences, as well as applying data mining to fault localization, is still in an early phase, and there are many avenues to explore and improve the effectiveness.

## 7.8. Fault Localization for GUI Software using N-gram Analysis[14]

After describing how to use *N*-gram analysis to conduct fault localization for program, let us see how *N*-gram analysis is used to localize the faults of GUI modules.

Unlike traditional software, test cases for testing GUI programs are event sequences instead of individual input data or files [20]. For each event, there must be a piece of source code that addresses the event; we call it the event's corresponding event handler [31]. By ranking these event handlers by their level of suspicion accounting for faults encountered during the testing, we can help programmers pinpoint the failures that cause event handlers in the source code. Due to the large volume of GUI event sequences, usually infinity, we apply data mining techniques to the test data to extract the most relevant sequences. From each test case, we generate *N*-grams, i.e., subsequences of length *N*. From these, we choose *N*-grams that appear more than a certain number of times in the event sequence of failing test cases. For these *N*-grams, we calculate the confidence: the conditional probability that a test case fails given that the *N*-gram appears in that test case's event sequence. We sort the *N*-grams in descending order of confidence and report the events of the program in the order of their first occurrence in the sorted list.

### 7.8.1. Background

#### 7.8.1.1. Representation of the GUI and its Operations

*GUI components*: Objects of a GUI include some windows and all kinds of GUI widgets (such as button, menu) in the windows [20]. In a broad definition, the window itself can also be viewed as a kind of GUI widget. Each window or GUI widget has a fixed set of properties, such as the size and position of the GUI widget. At any specific point in time, the GUI can be described in terms of the specific GUI widgets that it contains and the values of their properties.

*Event*: The basic inputs for GUI software are events. When a GUI application is running, users' operations trigger events, and the application responds to these events. Since events may be performed on different types of GUI widgets in different contexts, yielding different behaviors, they are parameterized with objects and property values. Commonly, we can use a 3-tuple <*a, o, v*> to represent event where *a* is the action of the event (such as clicking the mouse), *o* stands for the GUI widget the event deals with (such as a button), and *v* is the set of parameters of this action[31].

*GUI test cases*: Unlike traditional software, GUI test cases are usually defined as event sequences [20]. A formal representation of a GUI test case is a legal event sequence <$e_1, e_2, e_n$> consisting of *n* events. "Legal" refers to the fact that for $i = 1, 2, \cdots, n-1, e_{i+1}$ can be immediately accepted by GUI software and executed after the execution of $e_i$. Cai et al. extended this definition by defining GUI test cases using a hierarchical language [17]. People can first generate simple test cases as low order test cases. Then, more test cases can be generated by combining these low order test cases.

*EFG and EIG*: Due to the hierarchical relationship between GUI widgets, some events can only be accepted by the GUI software after the execution of some other events. If event $e_j$ can be immediately accepted and executed after the execution of event $e_i$, we say $e_j$ follows $e_i$ and ($e_i, e_j$) is called an event interaction. In model-based GUI testing, event-flow graph (EFG) is used to model all possible event interactions of GUI software [20]. EFG is a directed graph that contains nodes (that represent events) and edges (that represent a relationship between events).

---

[14] Part of Chapter 7.8 is from Reference [30]

Event-interaction graph (EIG) is an improvement for EFG [22]. EIG nodes, on the other hand, do not represent events to open or close menus or open windows. These events are only structural events and usually do not interact with the underlying software. The result is a more compact, and hence more efficient, GUI model.

Figure 7.10 presents a GUI that consists of four events: Undo, Redo, Select, and Edit. Figure 7.10 (b) shows the GUI's EFG; the four nodes represent the four events, while the edges represent the follows relationships. For example, in this EFG, the event Undo follows Edit, represented by a directed edge from the node labeled Edit to Undo. Figure 7.10 (c) shows the GUI's EIG. The Edit event is a menu open event, and we only use it to get the availability of the three other events, so the EIG model does not contain it. Following the deletion of event Edit, each edge ($e_x$, Edit) in the original EFG is replaced with edge ($e_x$, $e_y$) for each occurrence of edge (Edit, $e_y$).

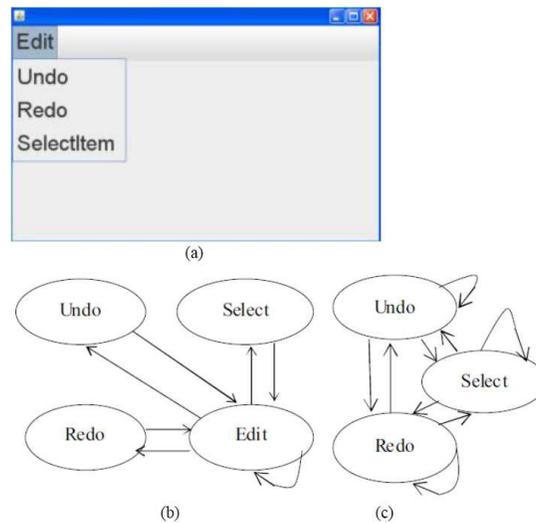

Figure 7.10. (a) A simple GUI, (b) its EFG, and (c) its EIG [30].

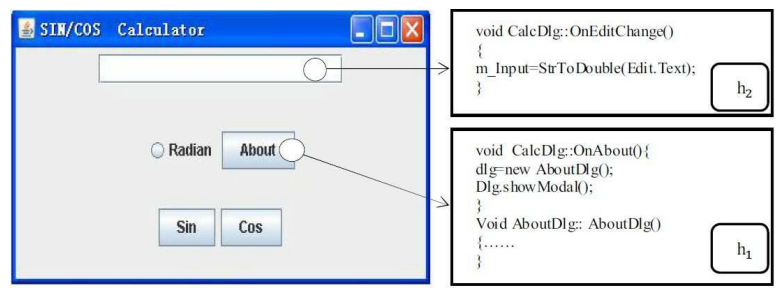

Figure 7.11. Example of event handler [30].

### 7.8.1.2. Event Handler

When one event executes on the software, all the pieces of source code that are possible to be executed are called the event's corresponding event handler [31]. The event handler usually includes the message response function and source code called by this function. Each event at most has one event handler. In addition, an event handler may response to several events. For example, pressing the shortcut key "Ctrl+S" has the same effect as clicking the "Save" button on the toolbar in Microsoft Word; therefore, these two events have the same event handler. Figure 8.12 gives an example of an event handler. The figure lists two events' event handlers out of five, $h_1$ and $h_2$, which respectively represent the event handler for the event of clicking the "About" button and the event of typing in the textbox.

### 7.8.1.3. N-gram

For a GUI application, let $T = \langle t_1, t_2, \cdots, t_n \rangle$ be the test set with n test cases in it. Each test case $t_i = \langle I_i, X_i \rangle$ is composed of the input $I_i$, i.e., a set of events and $X_i$, the expected output of $I_i$. When input $I_i$ is executed, it produces the actual output $A_i$. If $A_i = X_i$, then we regard $t_i$ as a passing test case, and if $A_i \neq X_i$; then $t_i$ is regarded as a failing test case. Let $E_i = \langle e_1, e_2 \cdots, e_k \rangle$ be the event sequence corresponding to test case $t_i$, then the total event sequences of the entire test set is $E = \langle E_1, E_2 \cdots, E_n \rangle$. We define two sets based on the outcome of the test cases passing event sequences which is $E_p = \{E_i | t_i \text{ is a passing test case}\}$ and failing event sequences which is $E_f = \{E_i | t_i \text{ is a failing test case}\}$.

Given an ordered list, an N - gram is any sub-list of N consecutive elements in the list. The elements of the N- gram must be in the same order as they were in the original list, and they must be consecutive. Given an event sequence $E_i$, an N-gram $G_{E_i,N,\alpha}$ is a contiguous subsequence $E_i = \langle e_\alpha, e_{\alpha+1}, \cdots, e_{N+\alpha-1} \rangle$ of length N starting at position $\alpha$. Taking the difference of $\alpha$ into account, the set of all N-grams for $E_i$ is $G_{E_i,N,1} = \{G_{E_i,N,1}, G_{E_i,N,2}, \cdots, G_{E_i,N,k-N+1}\}$. After deleting the duplicated N-grams, $G_{E,N} = \{G_{E_1,N}, G_{E_2,N}, \cdots, G_{E_n,N}\}$ is the N-grams corresponding to the total event sequences E.

### 7.8.2. Association Rule Mining

Association rule mining searches for interesting relationships among items in a given data set [11], which involves the following two procedures:

(1) *Frequent itemset generation*: Search for sets of items occurring together frequently, called a frequent itemset, whose frequency in the data set, called support, exceeds a predefined threshold (known as the *minimum* Support). For an N-gram $G_{E_i,N,p}$, Support is the number of failing test case containing $G_{E_i,N,p}$:

$$Support(G_{E_i,N,p}) = |\{E_j | G_{E_i,N,p} \in E_j \text{ and } E_j \in E_f\}|$$

(2) Association Rule Generation: Look for association rules like $A \Rightarrow B$ among the elements of the frequent itemsets, meaning that the appearance of A in a set implies the appearance of B in the same set. The conditional probability $P(B|A)$ is called Confidence, which must be greater than a predefined Minimum Confidence for a rule to be considered. For an N-gram $G_{E_i,N,p}$, Confidence is the conditional probability that the test case outcome is failure given that $G_{E_i,N,p}$ appears in the event sequence of that test case; That is:

$$Confidence(G_{E_i,N,p}) = \frac{|\{E_j | G_{E_i,N,p} \in E_j \text{ and } E_j \in E_f\}|}{|\{E_j | G_{E_i,N,p} \in E_j\}|}$$

### 7.8.3. Methodology

In this section, we present the methodology for GUI software fault localization [30]. As input, we use the source code, the test case types, and the event sequences of all the test cases, and we produce as output an ordered list of events, sorted in order of probability of their corresponding event handler containing the fault. We apply *N*-gram analysis to these event sequences to generate all possible unique *N*-grams for a given range of *N*. For each *N*-gram, we calculate its frequency in passing and failing event sequences. The set of *N*-grams and their frequencies are analyzed using the association rule mining technique described above.

The execution of the faulty event handler may not always cause failure of the test case. There may be quite a number of test cases in which the faulty event handler was executed but did not cause a failure. In most cases, the failure is dependent on the execution of other events. A specific event sequence of execution will cause the program to fail, and this event sequence will be very common in the failing event sequences but not so common in the passing event sequences. Therefore, we can find the subsequences that are most likely to contain the fault by analyzing the event sequences of passing and failing test cases.

#### A. General Approach

We model the events as items and the event sequence corresponding to a test case as the transaction. For example, $E_i = \langle e_1, e_2, , e_k \rangle$ is a transaction corresponding to the test case $t_i$. We generate frequent itemsets from the transactions with the additional constraint that the items in an itemset must be consecutive in the original transaction. To do this, we generate $G_{E,N}$, and from them, we choose the ones with at least the *minimum support*. Then, we try to discover association rules of the form $A \rightarrow failing$ from these itemsets where the antecedent is an *N*-gram and the consequent is failing. Therefore, the *N*-grams that appear as antecedents in the association rules are most likely to have caused the failure of the test case. We sort these *N*-grams in descending order of *confidence*. After sorting the *N*-grams, we convert the *N*-grams back to event orders.

#### B. N-Gram Fault Localization Algorithm

(1) *Parameters*: There are two major parameters in the algorithm: the first is *MinSup*, the minimum support for selecting the *N*-grams, and the second is $N_{MAX}$, the maximum value of *N* for generating the *N*-grams. Taking a low value of minimum support will result in the inclusion of irrelevant *N*-grams in consideration. Giving a high value of minimum support may cause us to discard some very important *N*-grams. Thus, we should give *MinSup* a median value. For GUI software, the execution of event $e_1$ may differ because of the execution of the previous event $e_2$. If event sequence <$e_1$, $e_2$> can be executed without inserting another intermediate event (such as an event to open menu), such differences can be detected by 2-grams; otherwise, such differences can be commonly detected by 3-grams. Consequently, we believe a choice of 3 for $N_{MAX}$ should give good results in most cases. If we give CDEF@higher value, the algorithm will still be able to find the fault, but due to larger N-grams, we will have to examine more event handlers to find the fault.

(2) Algorithm pseudo-code: In this section, the complete algorithm is presented in Algorithm 8.4.

(a) *Generate N-grams*: In this step, we first generate all possible *N*-grams of length 1 to CDEF from the event sequences of all test cases and then delete the duplicate ones. The generation of all $N_0$-grams from a set of event sequences for a given length $N_0$ is carried out in lines 2 to 5 of Algorithm 7.5, the deletion of the duplicate *CH*-grams is performed in lines 6 to 10 of

(b)
```
1 Function GenerateNGrams(E, N₀):
2     G ← φ
3     foreach Eᵢ ∈ E do
4         G ← G ∪ G_{Eᵢ,N₀}
5     end
6     foreach n ∈ G do
7         if n exists more than one time in G then
8             remove duplicate n from G
9         end
10    end
11    return G
```

(c) Algorithm 7.5, and the generation and combination of all the *N*-grams, *N* 1 to $N_{MAX}$, are done in lines 2 to 5 of

(d)
```
1 Function LocalizeFaults(E, E_f, MinSup):
2     NG ← φ for N = 1 to N_{MAX} do
3         NG ← NG ∪ GenerateNGrams(E, N)
4     end
5     E_{rel} ← {n|n ∈ NG and |n| = 1} foreach n ∈ E_{rel} do
6         if n do not exist in any event sequence of E_f then
7             Remove n from NG and E_{rel}
8         end
9     end
10    NG₁ ← {n|n ∈ NG and for all s ∈ E_{rel}, s ∉ n}
11    NG ← NG − NG₁
12    foreach n ∈ NG do
13        if Support(n) < MINSUP then
14            Remove n from NG
15        end
16    end
17    foreach n ∈ NG do
18        NF ← |{E_k|E_k ∈ E_F and n ∈ E_k}|
19        NT ← |{E_k|E_k ∈ E and n ∈ E_k}|
20        n.confidence ← NF ÷ NT
21    end
22    Sort NG in the descending order of confidence
23    Report the event numbers in the order of their first appearance in NG
```

(e) Algorithm 7.4.

(f) *Find relevant events*: From 1-gram, where the length of the sub-list is 1, we construct a set of relevant events, denoted by $E_{Rel}$. It contains only the events that have appeared at least in one of the failing event sequences in lines 6 to 11 of

(g)
```
 1  Function LocalizeFaults(E, E_f, MinSup):
 2    NG ← φ for N = 1 to N_MAX do
 3    │  NG ← NG ∪ GenerateNGrams(E, N)
 4    end
 5    E_rel ← {n | n ∈ NG and |n| = 1} foreach n ∈ E_rel do
 6    │  if n do not exist in any event sequence of E_f then
 7    │  │  Remove n from NG and E_rel
 8    │  end
 9    end
10    NG_1 ← {n | n ∈ NG and for all s ∈ E_rel, s ∉ n}
11    NG ← NG − NG_1
12    foreach n ∈ NG do
13    │  if Support(n) < MINSUP then
14    │  │  Remove n from NG
15    │  end
16    end
17    foreach n ∈ NG do
18    │  NF ← |{E_k | E_k ∈ E_F and n ∈ E_k}|
19    │  NT ← |{E_k | E_k ∈ E and n ∈ E_k}|
20    │  n.confidence ← NF ÷ NT
21    end
22    Sort NG in the descending order of confidence
23    Report the event numbers in the order of their first appearance in NG
```

(h) Algorithm 7.4.

```
1 Function LocalizeFaults(E, E_f, MinSup):
2     NG ← φ for N = 1 to N_MAX do
3         NG ← NG ∪ GenerateNGrams(E, N)
4     end
5     E_rel ← {n|n ∈ NG and |n| = 1} foreach n ∈ E_rel do
6         if n do not exist in any event sequence of E_f then
7             Remove n from NG and E_rel
8         end
9     end
10    NG_1 ← {n|n ∈ NG and for all s ∈ E_rel, s ∉ n}
11    NG ← NG − NG_1
12    foreach n ∈ NG do
13        if Support(n) < MINSUP then
14            Remove n from NG
15        end
16    end
17    foreach n ∈ NG do
18        NF ← |{E_k|E_k ∈ E_F and n ∈ E_k}|
19        NT ← |{E_k|E_k ∈ E and n ∈ E_k}|
20        n.confidence ← NF ÷ NT
21    end
22    Sort NG in the descending order of confidence
23    Report the event numbers in the order of their first appearance in NG
```

Algorithm 7.4. GUI fault localization using N-gram analysis [30]

```
1 Function GenerateNGrams(E, N_0):
2     G ← φ
3     foreach E_i ∈ E do
4         G ← G ∪ G_{E_i, N_0}
5     end
6     foreach n ∈ G do
7         if n exists more than one time in G then
8             remove duplicate n from G
9         end
10    end
11    return G
```

Algorithm 7.5. $N_0$-gram generation [30]

(i) *Eliminate irrelevant N-grams*: From lines 12 to 13 of

(j)
```
1  Function LocalizeFaults(E, E_f, MinSup):
2      NG ← φ for N = 1 to N_MAX do
3          NG ← NG ∪ GenerateNGrams(E, N)
4      end
5      E_rel ← {n|n ∈ NG and |n| = 1} foreach n ∈ E_rel do
6          if n do not exist in any event sequence of E_f then
7              Remove n from NG and E_rel
8          end
9      end
10     NG_1 ← {n|n ∈ NG and for all s ∈ E_rel, s ∉ n}
11     NG ← NG − NG_1
12     foreach n ∈ NG do
13         if Support(n) < MINSUP then
14             Remove n from NG
15         end
16     end
17     foreach n ∈ NG do
18         NF ← |{E_k|E_k ∈ E_F and n ∈ E_k}|
19         NT ← |{E_k|E_k ∈ E and n ∈ E_k}|
20         n.confidence ← NF ÷ NT
21     end
22     Sort NG in the descending order of confidence
23     Report the event numbers in the order of their first appearance in
       NG
```

(k) Algorithm 7.4, we discard the *N*-grams that do not contain any event from the relevant event set $E_{Rel}$.

(l) *Find frequent N-grams*: From lines 14 to 18 of

```
1  Function LocalizeFaults(E, E_f, MinSup):
2      NG ← φ for N = 1 to N_MAX do
3          NG ← NG ∪ GenerateNGrams(E, N)
4      end
5      E_rel ← {n|n ∈ NG and |n| = 1} foreach n ∈ E_rel do
6          if n do not exist in any event sequence of E_f then
7              Remove n from NG and E_rel
8          end
9      end
10     NG_1 ← {n|n ∈ NG and for all s ∈ E_rel, s ∉ n}
11     NG ← NG − NG_1
12     foreach n ∈ NG do
13         if Support(n) < MINSUP then
14             Remove n from NG
15         end
16     end
17     foreach n ∈ NG do
18         NF ← |{E_k|E_k ∈ E_F and n ∈ E_k}|
19         NT ← |{E_k|E_k ∈ E and n ∈ E_k}|
20         n.confidence ← NF ÷ NT
21     end
22     Sort NG in the descending order of confidence
23     Report the event numbers in the order of their first appearance in
       NG
```
(m)

(n) Algorithm 7.4, we eliminate *N*-grams with support less than the *MinSup*.

(o) *Rank N-grams by confidence*: For each surviving *N*-gram, we compute its confidence. This is done in lines 19 to 23 of

(p)

```
1  Function LocalizeFaults(E, E_f, MinSup):
2      NG ← φ for N = 1 to N_MAX do
3          NG ← NG ∪ GenerateNGrams(E, N)
4      end
5      E_rel ← {n | n ∈ NG and |n| = 1} foreach n ∈ E_rel do
6          if n do not exist in any event sequence of E_f then
7              Remove n from NG and E_rel
8          end
9      end
10     NG_1 ← {n | n ∈ NG and for all s ∈ E_rel, s ∉ n}
11     NG ← NG − NG_1
12     foreach n ∈ NG do
13         if Support(n) < MINSUP then
14             Remove n from NG
15         end
16     end
17     foreach n ∈ NG do
18         NF ← |{E_k | E_k ∈ E_F and n ∈ E_k}|
19         NT ← |{E_k | E_k ∈ E and n ∈ E_k}|
20         n.confidence ← NF ÷ NT
21     end
22     Sort NG in the descending order of confidence
23     Report the event numbers in the order of their first appearance in NG
```

(q) Algorithm 7.4. Then, we rank the *N*-grams in order of *confidence* in line 24 of

(r)

(s) Algorithm 7.4.

(t) *Convert N-grams to event numbers and rank events*: We traverse the ordered list of *N*-grams and report the event numbers in the order of their first appearance in the list. This is done in lines 25 to 26 of

```
1  Function LocalizeFaults(E, E_f, MinSup):
2      NG ← φ for N = 1 to N_MAX do
3          NG ← NG ∪ GenerateNGrams(E, N)
4      end
5      E_rel ← {n | n ∈ NG and |n| = 1} foreach n ∈ E_rel do
6          if n do not exist in any event sequence of E_f then
7              Remove n from NG and E_rel
8          end
9      end
10     NG_1 ← {n | n ∈ NG and for all s ∈ E_rel, s ∉ n}
11     NG ← NG − NG_1
12     foreach n ∈ NG do
13         if Support(n) < MINSUP then
14             Remove n from NG
15         end
16     end
17     foreach n ∈ NG do
18         NF ← |{E_k | E_k ∈ E_F and n ∈ E_k}|
19         NT ← |{E_k | E_k ∈ E and n ∈ E_k}|
20         n.confidence ← NF ÷ NT
21     end
22     Sort NG in the descending order of confidence
23     Report the event numbers in the order of their first appearance in NG
```

(u)

(v) Algorithm 7.4. If there are multiple *N*-grams with the same confidence, then concerning the order for "new events" (do not exist in *N*-grams with a higher confidence) that these *N*-grams contain, the best case will be the ordering in which the more "new faults" (do not exist in the event handler of higher ranking event) that one event handler contains, the earlier the position the corresponding event appears in the event group; we call this the best situation. The worst case will be the ordering in which the more "new faults" that one event handler contains, the later the position the corresponding event appears in the event group; we call this the worst situation. If two event handlers contain the same number of "new faults", we break the tie by random selection. For example, if we finally have two 2-grams <$e_1$, $e_2$> and <$e_3$, $e_4$> of the same confidence 0.8, suppose $e_1$, $e_2$, $e_3$, and $e_4$ are "new events" and the event handler individually contains 1,0,0,2 "new faults". Then, we obtain event orders "$e_4$, $e_1$, $e_2$, $e_3$" or "$e_4$, $e_1$, $e_3$, $e_2$" under the *best situation* and "$e_2$, $e_3$, $e_1$, $e_4$" or "$e_3$, $e_2$, $e_1$, $e_4$" under the *worst situation*.

### 7.8.4. Conclusion

Unlike traditional software, GUI test cases are usually event sequences, and each individual event has a unique corresponding event handler. This approach applies data mining techniques to the event sequences and their output data in terms of failure detections collected in the testing phase to rank the fault proneness of the event handlers for fault localization. This approach only focuses on how to prioritize the event handlers of the program under test. To further improve the efficiency, other fault localization techniques can be integrated with this approach so that code-level suggestions can be generated by analyzing the runtime execution data.